\documentclass[twocolumn]{aastex631}

\newcommand{\msun}{{\rm M}_\odot}
\newcommand{\zsun}{{\rm Z}_\odot}
\usepackage{newtxtext,newtxmath}

\usepackage[normalem]{ulem}

\title{Nitrogen pollution by metal enriched supermassive stars. }
\shortauthors{Nagele and Umeda}
\graphicspath{{./}{}}

\begin{document}

\title{Multiple channels for nitrogen pollution by metal enriched supermassive stars and implications for GN-z11. }

\correspondingauthor{Chris Nagele}
\email{chrisnagele.astro@gmail.com}
\author{Chris Nagele}
\affiliation{Department of Astronomy, School of Science, The University of Tokyo, 7-3-1 Hongo, Bunkyo, Tokyo 113-0033, Japan}
\author{Hideyuki Umeda}
\affiliation{Department of Astronomy, School of Science, The University of Tokyo, 7-3-1 Hongo, Bunkyo, Tokyo 113-0033, Japan}




\begin{abstract}

GN-z11 is an unusually luminous high redshift galaxy which was recently observed to have strong nitrogen lines while at the same time lacking traditional signatures of AGN activity. These observations have been interpreted as a super-solar nitrogen abundance which is challenging to explain with standard stellar evolution and supernovae enrichment. We present simulations of four models of metal enriched supermassive stars after the zero age main sequence which produce super-solar nitrogen consistent with the observations of GN-z11. We then show that the most massive model ends its life in a violent explosion which results in even greater nitrogen pollution. 

\end{abstract}



\section{Introduction}
\label{introduction}

Very massive stars (VMSs) and supermassive stars (SMSs) are hypothetical stars with masses above $100$ $\msun$, where the difference in nomenclature is used to indicate that SMSs collapse due to the general relativistic (GR) radial instability whereas the less massive VMSs collapse due to the pair instability \citep{fuller1986,woods2020}. Being so massive, these stars must be radiation dominated to maintain hydrostatic equilibrium \citep{Chandrasekhar1939isss.book.....C}, and thus have high entropy and luminosities near the Eddington limit \citep{shapiro1983}. V/SMSs are primarily studied in relation to the direct collapse scenario \citep[e.g.][]{woods2019,Haemmerle2020SSRv..216...48H,inayoshi2020} which aims to explain the observation of supermassive black holes at high redshift \citep{banados2018,wang2021,Eilers2022arXiv221116261E,Marshall2023arXiv230204795M}. This focus on the early universe, as well as theoretical constraints on the maximum metallicity at which direct collapse is possible \citep{Chon2020MNRAS.494.2851C,Hirano2022arXiv220903574H}, naturally leads to the study of metal free or extremely metal poor V/SMSs. However, metal enriched V/SMSs have been considered in two scenarios.

The first aims to explain the C–N and O–Na abundance anti-correlations in globular clusters \citep{Carretta2009A&A...505..139C,Carretta2009A&A...505..117C} which can be resolved by invoking hot CNO nucleosynthesis \citep{Prantzos2017A&A...608A..28P}. The conditions for this nucleosynthesis can be reached in a V/SMS formed by runaway collisions \citep{PortegiesZwart1999A&A...348..117P} and this scenario can explain the anti-correlations if enrichment ceases before much helium is produced \citep{Denissenkov2014MNRAS.437L..21D,Denissenkov2015MNRAS.448.3314D}. Furthermore, the observation of multiple stellar populations \citep{Piotto2015AJ....149...91P} is consistent with a V/SMS polluter if the star formation occurs during the lifetime of the V/SMS \citep{Gieles2018MNRAS.478.2461G}. 

\begin{figure*}
    \centering
    \includegraphics[width=2\columnwidth]{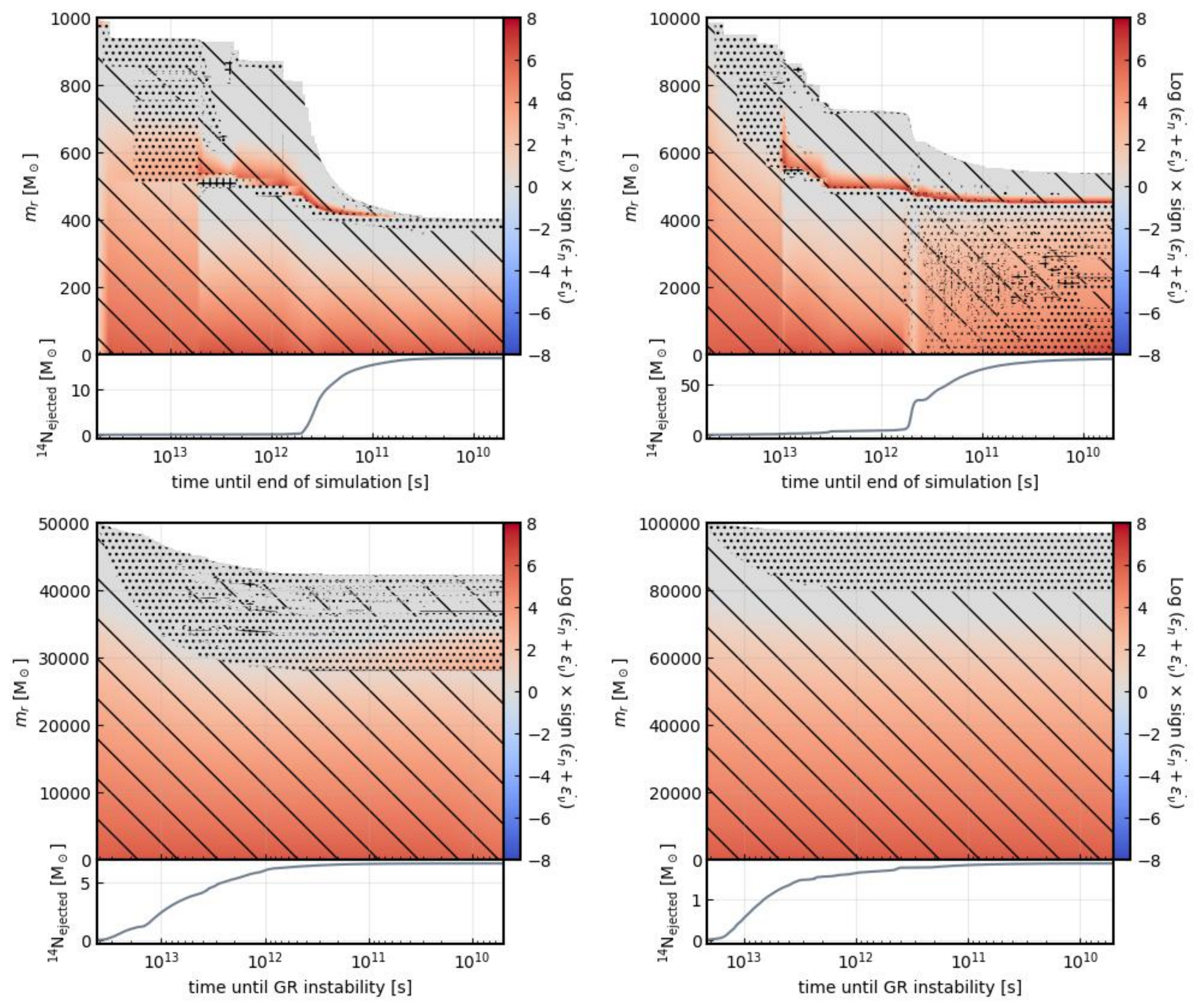}
    \caption{Kippenhahn diagrams for the four models. The horizontal axis is time until the end of the simulation (for the two lower mass models) or until the GR instability (for the two higher mass models) while the vertical axes are mass coordinate and cumulative $^{14}$N ejected from the stellar wind. In the upper half of each panel the color shows the heating and [very small] cooling rate from nuclear and neutrino reactions. The hatches show convective (diagonal), radiative (dotted) and semi-convective (crossed) regions.}
    \label{fig:evol}
\end{figure*}

\begin{figure}
    \centering
    \includegraphics[width=1\columnwidth]{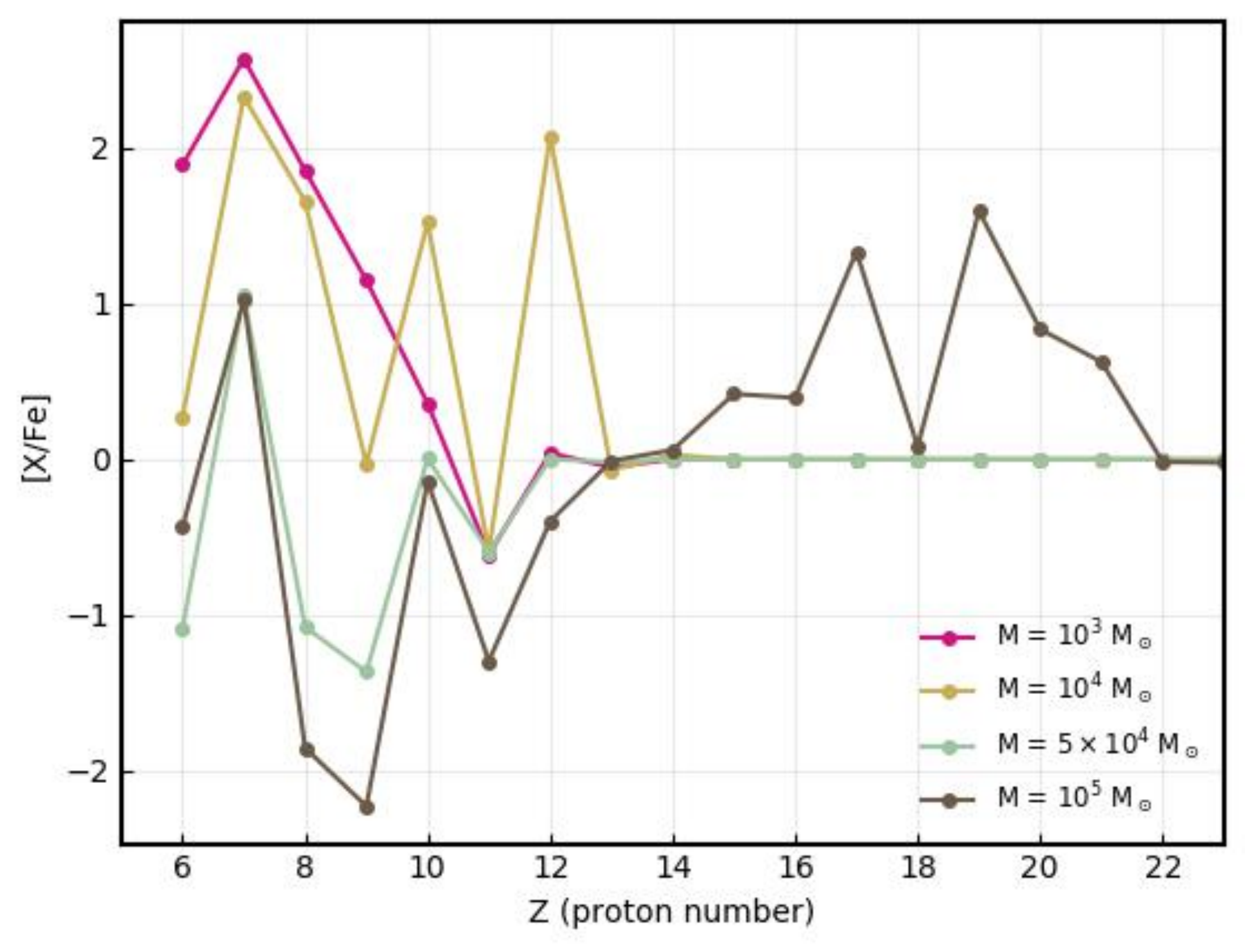}
    \caption{Log elemental yields, relative to Fe, relative to solar \citep{asplund2009}, for each of the four models. For the $10^5$ $\msun$ model, this includes not only wind driven mass loss, but also the explosion ejecta. }
    \label{fig:yields}
\end{figure}

\begin{figure*}
    \centering
    \includegraphics[width=2\columnwidth]{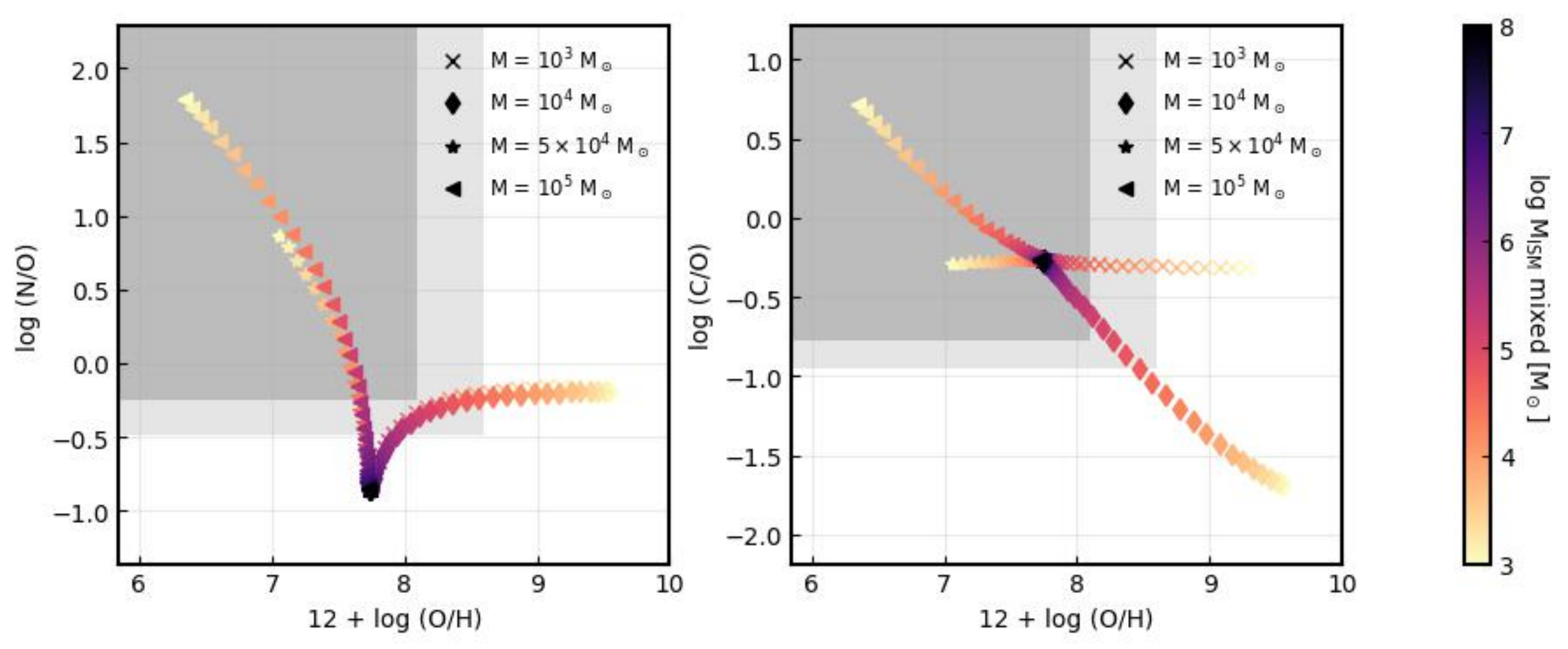}
    \caption{Abundance ratios of nitrogen to oxygen, carbon to oxygen, and oxygen to hydrogen. The symbols denote each of the four models, while the color shows how much Z$=0.1$ $\zsun$ ISM the yields from each model have mixed with. The gray regions show the conservative (lighter) and fiducial (darker) constraints on GN-z11 from the model of \citet{Cameron2023arXiv230210142C}. }
    \label{fig:CNOs}
\end{figure*}

The second scenario for metal enriched V/SMS formation involves the merger of two gas rich massive galaxies \citep{Mayer2010Natur.466.1082M,Mayer2019RPPh...82a6901M}. Gravitational torques funnel gas to small scales, where an optically thick rotationally supported disk is formed. When the disk becomes massive enough, it suffers gravitational collapse and forms a single central object, regardless of the gas metallicity \citep{Mayer2015ApJ...810...51M}. The key element in this scenario is the rapidity of the accretion which prevents cooling by metal lines and the resultant fragmentation. The central object is thought to be either a supermassive protostar with mass up to $10^9$ $\msun$ \citep{Haemmerle2019A&A...632L...2H,Haemmerle2021A&A...650A.204H} which collapses directly to a black hole or a V/SMS with mass $10^{3-5}$ $\msun$ which reaches the main sequence, depending on the virial masses of the merging galaxies \citep{Mayer2015ApJ...810...51M}. In a recent paper \citep{Nagele2023arXiv230101941N}, we showed that a fraction of metal enriched SMSs undergo an explosion powered by the CNO (carbon, nitrogen, oxygen) cycle and that the ejecta from these explosions is nitrogen rich. 

This brings us to GN-z11, a luminous galaxy slightly under redshift eleven with metallicity about ten percent the solar value \citep{Oesch2016ApJ...819..129O,Jiang2021NatAs...5..256J,Tacchella2023arXiv230207234T}. GN-z11 was recently observed by JWST NIRSpec to have unusually bright nitrogen lines, while simultaneously lacking the N V (E $>$ 77.5 eV) associated with an active galactic nucleus (AGN) \citep{Bunker2023arXiv230207256B}. \citet{Cameron2023arXiv230210142C} analyze the NIRSpec observations and infer a N/O ratio two to four times the solar value, though they also conclude that an AGN origin cannot be ruled out. They propose several explanations for this enhanced nitrogen, including mass loss from rotating WN stars, fine tuned metal poor supernovae (SNe) yields, tidal disruption by an intermediate mass black hole, or winds from well mixed V/SMSs. \citet{Senchyna2023arXiv230304179S} also analyze the NIRSpec observations and arrive at a similar conclusion regarding N/O. They then analyze spectra from nearby galaxies and suggest that GN-z11 could be a proto-globular cluster. \citet{Charbonnel2023arXiv230307955C} compare the CNO abundances in GN-z11 to nearby globular clusters, and further expand on the suggestion that the origin of the enhanced nitrogen could be from V/SMSs formed via stellar collisions. 

In this letter, we simulate four Z$=0.1\zsun$ SMS models with different masses (the least massive model is likely a VMS, but as it does not reach the pair instability in our simulations, we will refer to all four models as SMSs). We show that all models eject nitrogen via winds and the most massive model ejects even more nitrogen via an explosion. We then mix this ejecta with a Z$=0.1\zsun$ interstellar medium (ISM) and compare the results with the abundances inferred from GN-z11.

\section{Methods}
\label{methods}

In this section, we first outline the post Newtonian stellar evolution code HOSHI which we use to simulate the SMSs, and then we detail our evaluation of the general relativistic (GR) radial instability, GR hydrodynamics simulations of the explosion, and nucleosynthetic post processing of the explosion.

The HOSHI code \citep{takahashi2016,takahashi2018,takahashi2019,yoshida2019,nagele2020} is a 1D post Newtonian stellar evolution code which solves the stellar structure and hydrodynamical equations using a Henyey type implicit method. HOSHI includes a nuclear reaction network (52 isotopes), neutrino cooling, convection, mass loss, and rotation. The mass loss prescription is the same as in \citet{yoshida2019}, and we caution that these mass loss rates have huge uncertainties in the SMS regime, and may in reality be much stronger or weaker than in our simulations. We initiate the star in a high entropy, low temperature state with deuterium removed to expedite the arrival at the zero age main sequence. We then evolve the SMS until the model runs into numerical difficulties related to strong mass loss (late CO burning phase) or until the model becomes GR unstable.

We determine the stability of the SMS in HOSHI by solving the pulsation equation for a hydrostatic, spherically symmetric configuration in general relativity \citep{chandrasekhar1964,haemmerle2020,Nagele2022arXiv220510493N}.The star is unstable if there exists a perturbation $\xi(r) \propto e^{i\omega t}$ with $\omega^2 < 0$ which will grow exponentially. We solve for $\omega_0^2$ by computing $\xi_0$ for several guesses of $\omega_0^2$ and then extrapolating the Wronskian of these solutions to find a more accurate value of $\omega_0^2$ (see appendices in \citet{Nagele2022arXiv220510493N} or \citealt{Nagele2023arXiv230101941N}). Once an unstable HOSHI profile is found, we port that profile to a 1D Lagrangian GR hydrodynamics code which uses a Roe-type approximate linearized Riemann solver \citep{yamada1997,takahashi2016,nagele2020}. This code includes all of the physics from HOSHI except for convection and ionization. We use a 153 isotope network which covers the proton rich side (p side) at a depth of 3-8 isotopes up to zinc. We then post process this model using a deeper 514 isotope network \citep{Nagele2023arXiv230101941N}. Reaction rates for all networks are taken from \textit{JINA REACLIB} \citep[][]{Cyburt2010ApJS..189..240C}.

\section{Results}
\label{results}

\subsection{Stellar Evolution}

All four of the models in this letter are almost completely convective at the start of hydrogen burning (Fig. \ref{fig:evol}). Radiative surface layers develop during late hydrogen burning. During this phase, the $10^5$ $\msun$ model becomes GR unstable. Just before hydrogen depletion, shell burning begins in the radiative region, soon after which the $5\times 10^5$ $\msun$ becomes GR unstable, with a central hydrogen mass fraction of $\sim 10^{-6}$. The two low mass models survive the transition to helium burning and eventually form carbon-oxygen cores. During this stage, the hydrogen shell burning intensifies, resulting in a largely convective envelope. In addition, helium shell burning begins in a thin layer just below the hydrogen shell burning region. During this phase, dredge up of helium, carbon, and oxygen from the core is efficient, eventually resulting in super-solar abundances of carbon and oxygen. The elemental abundances from the mass loss due to winds (and the explosion, for the $10^5$ $\msun$ model) is shown in Fig. \ref{fig:yields} and selected isotopes are listed in Table \ref{tab:yields}.

As we will show, this super-solar oxygen is not seen in GN-z11 so it is worth pausing to make a general comment about wind compositions. SMS winds in the central hydrogen burning phase will be enriched by the CNO cycle (both central burning and shell burning), so that they contain super-solar nitrogen and sub-solar carbon and oxygen. However, during the central helium burning phase, carbon and oxygen begin to accumulate in the core, some of which is dredged up to the envelope. During this phase, the wind consists of super-solar nitrogen and roughly solar carbon and oxygen abundances. Finally, after helium burning, dredge up from the CO core continues and the wind becomes super-solar in C and O. While the exact details of the cumulative wind composition will depend on mass, metallicity and rotation, the end of life burning phase can be used as a guidepost. 

\subsection{Hydrodynamics}

After the $5\times 10^5$ $\msun$ and $10^5$ $\msun$ models become GR unstable, we port the unstable stellar profiles into the hydrodynamics code to determine if they will explode or collapse. $5\times 10^5$ $\msun$ becomes unstable just at the end of hydrogen burning, and therefore it does not have enough hydrogen to power a CNO driven thermonuclear explosion \citep[][]{fuller1986,montero2012,Nagele2023arXiv230101941N} nor enough oxygen ($X \sim 10^{-4}$) to power an alpha process driven thermonuclear explosion \citep{chen2014,nagele2020,Nagele2022arXiv220510493N} and therefore this model collapses. On the other hand, the $10^5$ $\msun$ model has a healthy hydrogen reservoir and explodes with total energy $7.2\times 10^{54}$ ergs at shock breakout and a maximum outflow velocity of $9.8\times 10^8$ cm/s. 

After completing the hydrodynamical simulation, we post process the nucleosynthesis for the exploding model using a 514 isotope network. This step is not critical for this particular model, as the central temperature never goes above $3.4\times 10^8$ K. However, these temperatures are large enough to create some p-side isotopes such as $^{27-28}$P which are not included in the 153 isotope network. Besides the CNO cycle, this explosion is characterized by transport of light elements to slightly higher mass (proton number 15-20) by repeated proton captures. In this case, the ejecta has super-solar abundances of chlorine, potassium, calcium and scandium (Fig. \ref{fig:yields}). Similar models which reach higher temperatures have even higher ratios for these elements and more massive elements extending all the way to molybdenum \citep{Nagele2023arXiv230101941N}, and thus the identification of lines associated with these elements would be a strong piece of evidence for a SMS explosion. 

\begin{table*}
	\centering
	\caption{Hydrogen, helium, CNO, and magnesium yields in units of $\msun$ for each model.}
	\label{tab:yields}
	\begin{tabular}{crrrrrr} 
		    \hline \hline
    		 & $^{1}$H [M$_\odot$] & $^{4}$He [M$_\odot$] & $^{12}$C [M$_\odot$] & $^{14}$N [M$_\odot$] & $^{16}$O [M$_\odot$]  & $^{24}$Mg [M$_\odot$]   \\
    		\hline
 $10^3$ M$_\odot$  &  205  &  334  &  9.91  &  17  &  27  &  0.0368  \\
 $10^4$ M$_\odot$  &  1.65e3  &  2.73e3  &  1.81  &  75.8  &  133  &  31.2  \\
 $5\times 10^4$ M$_\odot$  &  5.06e3  &  2.7e3  &  0.142  &  6.69  &  0.412  &  0.441  \\
 $10^5$ M$_\odot$  &  4.19e4  &  5.8e4  &  5.63  &  81.4  &  0.884  &  1.42  \\
		\hline \hline
	\end{tabular}
\end{table*}

\subsection{Implications for GN-z11}

Fig. \ref{fig:CNOs} shows the abundance ratios of nitrogen to oxygen, carbon to oxygen, and oxygen to hydrogen as in Fig. 1 of \citet{Cameron2023arXiv230210142C} and Fig. 1 of \citet{Charbonnel2023arXiv230307955C}. The grey regions are the conservative and fiducial constraints derived by \citet{Cameron2023arXiv230210142C} (Table 1), where we have filled in the fiducial value of $12 + \rm log(O/H) \approx 8.1$ by inspection. The symbols show the yields from our models mixed with different masses of Z$=0.1\zsun$ gas as denoted by color. The yields from the lower mass models ($10^3,10^4$ $\msun$) are nitrogen rich, but also contain super-solar oxygen. These models survive into carbon-oxygen burning, at which point dredge up increases the oxygen and carbon abundances in the envelope. The implication of the oxygen rich winds is that although these two models produce super-solar nitrogen, they do not satisfy the fiducial constraints (dark grey region) and they only satisfy the conservative constraints (light grey region) for a narrow range of mixed ISM ($\rm M_{\rm ISM} \in (8e3,9e4)$ $\msun$ and $\rm M_{\rm ISM} \in (4e4,5e5)$ $\msun$ for the $10^3,10^4$ $\msun$ models, respectively).

The two higher mass models in this study both end their lives before helium burning, thus producing super-solar nitrogen and sub-solar oxygen. The models both satisfy the fiducial constraints for a wide mass range of mixed ISM ($\rm M_{\rm ISM}<4e4$ $\msun$ and $\rm M_{\rm ISM}<5e5$ $\msun$ for the $5\times 10^4,10^5$ $\msun$ models, respectively). We note that these ISM masses are small compared to the stellar mass of the galaxy ($\sim 10^9$ $\msun$), and if the galaxy has a homogeneous distribution of nitrogen, that would rule out nitrogen enrichment by a single V/SMS. This issue could be resolved if the nitrogen has an inhomogeneous distribution (e.g. concentrated near the site of the V/SMS) or if multiple V/SMSs were present (as has been suggested by other authors: \citealt{Cameron2023arXiv230210142C,Charbonnel2023arXiv230307955C}). This problem highlights one of the strengths of the V/SMS enrichment scenario, namely the large mass of nitrogen produced (Table \ref{tab:yields}). Other exotic enrichment scenarios (e.g. tidal disruption, peculiar supernovae yields) which do not produce as much nitrogen per event will have to have occurred many times in order for nitrogen to be observed at super-solar abundances.
 
 The two high mass models may overproduce nitrogen relative to oxygen and veer out of the fiducial constraints on the left side of the grey region, the edge of which is not shown here (see Fig. 1 of \citealt{Cameron2023arXiv230210142C}). That being said, this depends heavily on the pre-SMS oxygen abundance in the ISM, a quantity which could feasibly span several orders of magnitude, and we argue that these two models naturally fall within the fiducial constraints, as their primary nucleosynthetic characteristic is nitrogen production. More generally, super-solar nitrogen abundances with roughly solar oxygen abundances can be achieved by supermassive stars with intermediate end times, and the inclusion of additional intermediate mass models would fill in the region between the high mass and low mass curves in Fig. \ref{fig:CNOs}.

\section{Discussion}
\label{discussion}

In this letter, we have presented four models of metal enriched supermassive stars and showed that super-solar nitrogen pollution may originate both from stellar winds and explosions. In addition, there is the prospect that a set of models with better mass resolution will find supermassive star pulsations, either at the end of hydrogen burning or the end of helium burning \citep{Nagele2022arXiv220510493N,Nagele2022arXiv221008662N}. These pulsations would eject similar composition to the stellar winds, but in bulk quantities up to several thousands of solar masses. Since these pulsations would occur before the carbon-oxygen burning phase, they would have the CNO ratios required to explain GN-z11.

In our discussion thus far, we have only considered mass loss during and after the main sequence, but mass loss during the accreting phase could be just as or more important. This is due to the possibility of a conveyor belt effect, where mass is lost to winds, while simultaneously being replenished by gas accretion and stellar collisions \citep{Gieles2018MNRAS.478.2461G}. Supermassive stars with this conveyor belt mechanism, and which are massive enough, not only can explain the enhanced nitrogen in GN-z11, but can also explain the anti-correlations found in globular clusters. When and if accreting SMSs transition to non accreting SMSs is still largely unknown, although UV feedback during hydrogen burning is thought to be able to drive the transition, particularly in the case of sporadic accretion \citep{Sakurai2015MNRAS.452..755S}. 

For both the accreting and non-accreting supermassive stars, the salient question is when do they collapse? If collapse does not occur before carbon burning, then both the globular cluster abundances and the GN-z11 observations become much more difficult to reproduce. In addition, if the collapse occurs in the middle of helium burning, it is likely to trigger a thermonuclear explosion rich in alpha process elements \citep{Nagele2022arXiv220510493N} which would also rule out the SMS polluter theory. Thus, collapse either in hydrogen burning, or in very late helium burning is required and this may favor the high masses predicted by the merger scenario \citep{Mayer2010Natur.466.1082M}. Tantalizingly, the NIRCam observations of GN-z11 found nearby haze which could be evidence of a past merger \citep{Tacchella2023arXiv230207234T}. 

Ultimately, more precise observations are required in order to determine the veracity and origin of the enhanced nitrogen in GN-z11. In particular, better CNO abundances would help to constrain various SMS models as well as other exotic theories. Furthermore, an observation of enhanced scandium or vanadium would be strong evidence for an SMS explosion (unfortunately, upper limits would not rule out an explosion if the central temperature was relatively low, see \citealt{Nagele2023arXiv230101941N}). In addition, imaging followups could confirm the compactness of GN-z11 and the existence of the haze. This haze may in fact be a supernova remnant, as material from SMS explosions in this energy range are thought to reach several hundred parsecs before falling back into the halo and igniting a violent starburst \citep{Whalen2013ApJ...777...99W,Whalen2013ApJ...774...64W}, such as the one observed in GN-z11. Whatever this exciting galaxy turns out to be, it will surely advance our understanding of the conditions in the early universe.

\section*{Data Availability}

The data underlying this article will be shared on reasonable request to the corresponding author.

\section*{Acknowledgements}

This study was supported in part by the Grant-in-Aid for the Scientific Research of Japan Society for the Promotion of Science (JSPS, No. JP21H01123).


\bibliography{bib}{}
\bibliographystyle{aasjournal}



\end{document}